\documentclass[pre,preprint,aps,superscriptaddress,showpacs,floatfix]{revtex4}
\usepackage{graphicx}
\usepackage{amsmath}
\begin{document}
\title{Anticipated synchronization in coupled inertia ratchets with
  time-delayed feedback: a numerical study}

\author{Marcin Kostur}
\affiliation{Institut f\"ur Physik, Universit\"at Augsburg,
Universitatsstrasse 1, D-86135 Augsburg, Germany}

\author{Peter H\"anggi}
\affiliation{Institut f\"ur Physik, Universit\"at Augsburg,
Universitatsstrasse 1, D-86135 Augsburg, Germany}

\author{Peter Talkner}
\affiliation{Institut f\"ur Physik, Universit\"at Augsburg,
Universitatsstrasse 1, D-86135 Augsburg, Germany}

\author{Jos\'e L. Mateos}
\affiliation{Instituto de F\'{\i}sica,
Universidad Nacional Aut\'onoma de M\'exico, Apartado Postal 20-364,
01000 M\'exico, D.F., M\'exico}

\date{\today}
\begin{abstract}
  We investigate anticipated synchronization between two periodically
  driven deterministic, dissipative inertia ratchets that are able to
  exhibit directed transport with a finite velocity.  The two ratchets
  interact through an unidirectional delay coupling: one is acting as
  a master system while the other one represents the slave system.
  Each of the two dissipative deterministic ratchets is driven
  externally by a common periodic force.  The delay coupling involves
  two parameters: the coupling strength and the (positive-valued)
  delay time. We study the synchronization features for the unbounded,
  current carrying trajectories of the master and the slave,
  respectively, for four different strengths of the driving amplitude.
  These in turn characterize differing phase space dynamics of the
  transporting ratchet dynamics: regular, intermittent and a chaotic
  transport regime.  We find that the slave ratchet can respond in
  exactly the same way as the master will respond in the future,
  thereby anticipating the nonlinear directed transport.
\end{abstract}

\vspace{0.8cm}
\maketitle

\section{Introduction}
The intriguing concept of synchronization in nonlinear systems is
relevant for a wide range of topics in physics, chemistry and biology.
Recently, it has received much attention and first, comprehensive
reviews and books have appeared \cite{piko,bocca}. The case of
synchronization, in particular of chaotic systems, represents a
challenge, since a chaotic system is extremely sensitive to small
perturbations. Nevertheless, it has been established repeatedly that
the synchronization of chaotic systems is possible under certain
conditions \cite{chaos}.

The situation when the \textit{coupling} involves a delay in time is
the focus here, and may lead to  anticipated synchronization
\cite{voss1,voss2,voss3,voss4,ciszak,toral}.  In this case, one deals
with two systems: a ``master'' and a ``slave'', which are coupled
unidirectionally via a time-delay term, in such a way that, under some
circumstances, the slave system anticipates the response of the master
system. The regime of anticipated synchronization and its stability
has been studied theoretically previously for a variety of systems: we
mention here the case of linear set ups with delay \cite{calvo},
coupled chaotic maps with delays \cite{zanette,hernandez}, excitable
systems \cite{ciszak2} and nonlinear systems of practical
interest, such as semiconductor lasers operating in a chaotic regime
\cite{masoller}. Recently, the phenomenon of anticipated
synchronization has been vindicated experimentally in a semiconductor
laser running in the chaotic regime \cite{siva,tang}.

In different context, we witness an increasing interest during recent
years in the study of intriguing transport phenomena of nonlinear
systems that can extract usable work from unbiased non-equilibrium
fluctuations.  These, so called Brownian motor systems (or Brownian
ratchets) \cite{BM} can be modelled by a Brownian particle undergoing
a random walk in a periodic asymmetric potential, and being acted upon
by an external time-dependent force of zero average. The recent burst
of work is motivated by both, (i) the challenge to model
unidirectional transport of molecular motors within the biological
realm and, (ii) the potential for novel technological applications
that enable an efficient scheme to shuttle, separate and pump 
particles on the micro- and even nanometer scale \cite{BM}.

Although the vast majority of the literature in this field considers
the presence of noise, there have been attempts  to model the
transport properties of classical deterministic inertial ratchets
\cite{jun,ma1,ma2}. These ratchets generally possess parameter regions
where the classical dynamics is chaotic; the latter  in turn
then decisively determines the transport properties.

In contrast to the case with two coupled oscillators possessing
confined position trajectories, the position in a driven ratchet
dynamics is able to undergo directed, unbounded motion with a finite
transport velocity. In the following, we shall explore the case of a
coupling between two deterministic, dissipative ratchets in parameter
regimes where each ratchet system is individually able to exhibit
either current-carrying, regular transport trajectories (which in turn
possess a periodic velocity), or also a chaotic or even an
intermittent, directed transport behavior \cite{jun,ma1,ma2}.
One of the ratchet devices then acts as the ``master system'' while
the remaining one acts as the ``slave''.
The coupling between the two ratchets is unidirectional,
meaning that the master affects the slave, but not vice versa. We
shall explore the possibility of anticipated synchronization in various
tailored parameter regimes of a regular, chaotic and intermittent dynamics.

\section{Two coupled inertial ratchets with a time delay}

To start out, let us consider a one-dimensional problem of an inertial
particle driven by a periodic time-dependent external force in an
asymmetric (with respect to the reflection symmetry) periodic,
so called ratchet potential.  In order to have no net bias at work
the time average of the external force equals zero. Here, we do not
take into account any sort of noise, meaning that the dynamics is
deterministic. We thus deal with a rocked deterministic ratchet
\cite{jun,bartussek94} that obeys the following dimensionless inertial
dynamics \cite{ma1}:

\begin{equation}
\ddot{x} + b\dot{x} + {\frac{dU(x)}{dx}} = a\cos (\omega t),
\end{equation}

\noindent where $b$ denotes the friction coefficient,
$V(x)$ is the asymmetric ratchet periodic potential, $a$ is the
amplitude of the external force and $\omega$ is the frequency of the
external driving force. The dimensionless potential (which is shifted
by an amount $x_{0}\simeq -0.19$ in order that its minimum is located
at the origin) is given by:
\begin{equation}
U(x) = C - U_{0} \left [\sin 2\pi (x-x_{0}) + {\frac{1}{4}} \sin 4\pi
(x-x_{0}) \right ]
\end{equation}
\noindent and is depicted in Fig. 1. The constant $C$ is chosen
such that $U(0)=0$, and is given by $C = -U_{0} (\sin 2\pi x_{0} +
0.25 \sin 4\pi x_{0})$, where $U_{0} =1/4\pi^2(\sin (2\pi |x_{0}|) +
\sin (4\pi |x_{0}|))$.  In this case  $U_{0}
\simeq 0.0158$ and $C\simeq 0.0173$, see also  \cite{bartussek94,ma1}.

\begin{figure}[htb]
\centerline{\includegraphics{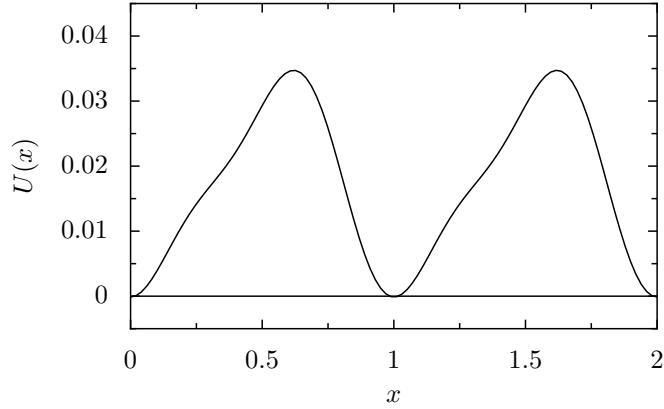}}
\caption{The dimensionless periodic ratchet potential used in our simulations.}
\label{fig1}
\end{figure}

This time-dependent dynamics can be embedded into a corresponding
three-dimensional autonomous phase space dynamics because we are
dealing with a Newtonian dynamics subjected to time dependent harmonic
driving.  Consequently, the equation of motion of the ratchet system
can be recast as a three-dimensional dynamical system.

We next consider two ratchets, coupled in a master-slave configuration
with $X$ denoting the coordinate of the master and $x$ that of the
slave. The positive valued delay coupling $K$ is one-way, that is, the
slave is coupled to the master, but the latter is independent of the
dynamics of the slave.  The two ratchets obey the coupled system of
equations:

\begin{equation}
\ddot{X}(t) + b\dot{X}(t) + U'(X(t)) = a\cos (\omega t),
\end{equation}

\begin{equation}
\ddot{x}(t) + b\dot{x}(t) + U'(x(t)) = a\cos (\omega t) + K(\dot{X}(t) - \dot{x}(t-\tau))\;,
\end{equation}

\noindent with $\tau$ denoting the (positive-valued) time-delay.  We
can mathematically recast this dynamical system in terms of two
coupled, autonomous three-dimensional dynamical systems, i.e.,

\begin{eqnarray}
\label{eq:system1}
\dot X(t) &=& V(t) \nonumber \\
\dot V(t) &=& -b V(t) - U'(X(t))+ a \cos(\Phi(t)) \nonumber \\
\dot \Phi(t) &=& \omega \nonumber \\
\dot x(t) &=& v(t) + K (X(t)-x(t-\tau)) \nonumber\\
\dot v(t) &=& -b v(t) -U'(x(t))+ a \cos(\varphi(t))  \nonumber \\
\dot \varphi(t) &=& \omega
\end{eqnarray}

Given this system, we notice that the manifold $X(t) = x(t-\tau)$
presents an exact solution of the system, when the period of the
external force is equal to the time delay $\tau$.  The situation when
we attain anticipated synchronization yields $X(t) = x(t - \tau)$, or
$X(t + \tau) = x(t)$. This implies that the slave ratchet acts at time
$t$ in exactly the same way as the master will do in the future
time, $t+\tau$, thus anticipating the dynamics.

We note, that the phase difference $\Phi(t)-\varphi(t)$ of the driving
forces remains fixed during its time evolution. Thus, the only way to
achieve the synchronization -- with anticipation time $\tau$ -- is to
choose the initial phase $ \varphi(0)$ of the slave to match precisely
\begin{equation}
\Phi(0) = \varphi(0)+\omega \tau.\label{eq:phi0}
\end{equation}

This dynamical system is similar to the excitable system studied in
\cite{ciszak2}, where the Adler equation \cite{adler} is considered for a particle
in a tilted symmetric (non-ratchet) periodic potential. In our case the number
of degrees of freedom is increased since we are addressing the inertial
dynamics.  We remark, however, that in the limit of two over-damped ratchets,
our system dynamics becomes  similar to the one studied previously
in \cite{ciszak2}, except that we deal here with a common periodic
forcing instead of a common random external forcing used in \cite{ciszak2}.

\section{Anticipated synchronization: numerical results}

\begin{figure}[htb]
\centerline{\includegraphics{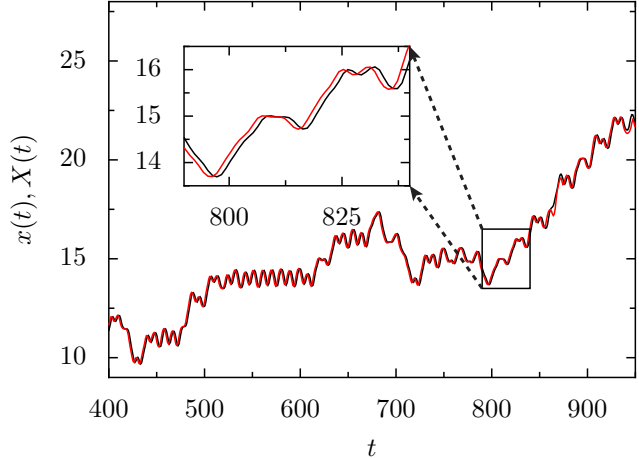}}
\caption{(Color online)
  Trajectories for the master (red line) and slave (black line) ratchets in a typical
  synchronization scenario. The magnified part of the trajectory
  reveals the anticipation effect.  The parameters are: $a = 0.08$, $b
  = 0.1$, $\omega= 0.67$, $K = 0.6$, $\tau = 1.3$.
}
\label{fig2}
\end{figure}

In the following, we numerically analyze thoroughly the dynamics of
the two coupled ratchets (\ref{eq:system1}).  Let us consider the case
where both ratchets, master and slave, are identical, that is, they
have the same parameters $a, b$, and $\omega$; the parameters that
enter in the ratchet potential are also identical.  In Fig. 2 we
depict transporting trajectories for the master and the slave ratchet,
when $a = 0.08$, $b = 0.1$ and $\omega = 0.67$. The delay coupling
involves two parameters: the coupling strength $K=0.6$ and the delay
time $\tau=1.3$.  We notice that the master and the time-shifted slave
essentially coincide; that is, the slave is anticipating the response
of the master system, see in the blow up in Fig. \ref{fig2}.

The parameter space of the full nonlinear dynamical system is far too
large for a systematic numerical analysis. In this paper we will
choose the dynamics of the master to be in one of the following
representative regimes: regular, chaotic and intermittent transport
dynamics.  Then, the relevant parameter space for synchronization
$(\tau,K)$ will be scanned numerically. We also seek a quantity that
can characterize the quality of synchronization.  A first natural
candidate would be the position correlation function. We found,
however, that this measure does not provide an intuitive answer
whether the master and the slave are synchronized. Instead, we will
use the fraction of the time during which the two ratchets are
synchronized within some prescribed accuracy.  That is, we consider
that the master and the slave attain a regime of anticipated
synchronization when the difference between their trajectories is
smaller than some small given value $\epsilon$, that is,
$|x(t)-X(t+\tau)| \le \epsilon$.  We always have set this value to
read $\epsilon=0.01$.

In the following, we calculate the trajectories for both ratchet
dynamics, compute the amount of time that they stay synchronized
(according to the above criterion), and then determine the ratio $p$
between the synchronized and total timespan of the considered
trajectory.  This measure of synchronization $p$ therefore varies
between zero and one.

\begin{figure}[htb]
\centerline{\includegraphics[width=8cm]{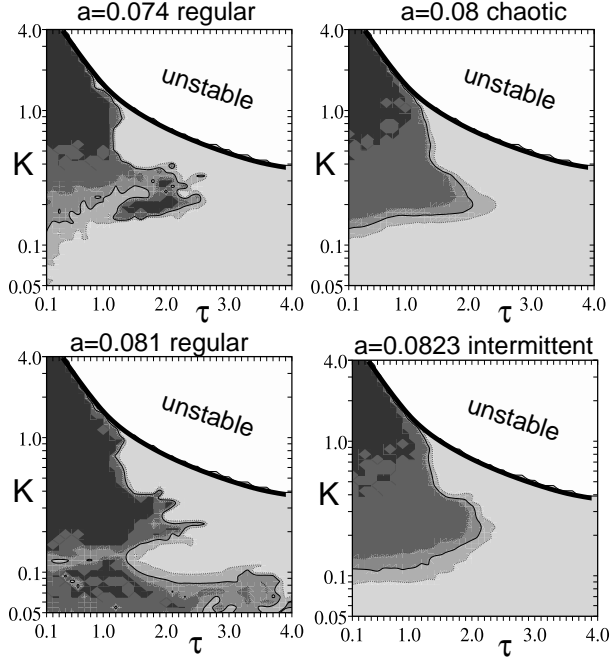}}
\caption{ Parameter space indicating the stability region for the
  anticipated synchronization to take place. We depict four cases: one
  with intermittent, directed transport dynamics at a driving
  amplitude of $a=0.0823$, one with a chaotic transport behavior at
  $a=0.08$ and two regular situations with $a=0.074$ and $a=0.081$,
  respectively.  The remaining parameters are as follows: $b =0.1$,
  $\omega = 0.67$.  Within the white area the slave exhibits
  exponentially growing oscillations. The gray-scale represents the
  quality parameter $p$ (defined in the text).  Four gray-tones from
  light to dark gray correspond to the corresponding ranges of $p$:
  $0\dots 0.25$, $0.25\dots 0.5$, $0.5\dots 0.75$ and $0.75\dots 1$,
  respectively. 
  Each point in this plot was obtained by averaging the ratio $p$ over
  $40$ trajectories of the system (\ref{eq:system1}) with random initial
  conditions. Each trajectory has a length of $20000$ time units.
  Every plot consumed ca.  200 hours of CPU on a typical
  workstation. The bold line exhibits the necessary stability
  condition $K \tau = \pi /2$ resulting from the linear part of
  the delay equation, see also the text.
}
\label{fig3}
\end{figure}

>From the literature it is known \cite{voss1,voss2,toral} that the
delay coupling scheme used here requires some constraints on the
positive anticipation time $\tau$ and the coupling strength $K$ for
the anticipated synchronization to occur. Thus, we investigate the
stability regions, in the parameter space $(K,\tau)$, for coupled
chaotic ratchets as described by eq. (\ref{eq:system1}).  In Fig.
\ref{fig3} we show the synchronization properties for the case $b =
0.1$ and $\omega = 0.67$, for four values of the amplitude $a$. These
values, indicated in the figure, correspond to regular, intermittent
and a chaotic regime of the master dynamics \cite{ma1}.  The
gray-scale depicts the value of the fraction of synchronization time
$p$, as defined above.  The darkest region corresponds to a well
synchronized behavior with values of $p$ being close to $1$; the light
gray represents an unsynchronized master and slave behavior with a
value $p\simeq 0$.


In all four panels in Fig. \ref{fig3} we can observe regions of
synchronization. Although, their shapes differ from panel to panel
they contain common features. First of all, we note that at
sufficiently small $K$, and in particular when $K=0$, the slave and
the master are not coupled and evolve independently of each other.
Secondly, the increase of the coupling strength $K$ causes the onset
of synchronization for small and moderate values of the delay time
$\tau$.  Synchronization is lost, however, for too large values of the
delay time $\tau$ regardless of the coupling $K$.  The third common
feature is the loss of the stability of the slave, if both $K$ and
$\tau$ are too large. The origin of the instability derives from the
linear delay equation $\dot x(t)=-Kx(t-\tau)$, resulting from
eq.~(\ref{eq:system1}) upon neglecting the velocity $v(t)$, see also
Ref.~\cite{calvo}. For $K \cdot \tau>\pi/2$ the trajectories
of this linear equation grow exponentially.  This criterion agrees
perfectly well with our numerical findings. The unstable region can
be neighboring to the synchronized or to unsynchronized one, depending
on the delay $\tau$.  Hence, two scenarios have been observed:
increasing $K$ at larger $\tau$ first de-synchronizes the system until
the slave finally becomes unstable (see e.g. $a=0.08$, $\tau=1.7$ and
$K=0.2\dots 4$); at smaller values of $\tau$ the synchronized state
becomes unstable with growing $K$ (see e.g.  $a=0.08$, $\tau=1.0$ and
$K=0.2\dots4$).

One can distinguish two regions of synchronization in these four plots
in Fig. \ref{fig3}.  The first one is present in all cases for
coupling strengths $K>0.1$.  It has a similar shape although the
underlying dynamics may differ dramatically.  The second region for
$K<0.1$ exists in the case of a regular dynamics with $a=0.081$.
Remarkably, in this regime the system can reach synchronization for
delay times $\tau$ much larger than in the other cases, and it takes
place even at a small coupling strength $K\simeq0.06$.

Clearly, the parameter $p$ does not provide the full information on
the dynamics. If we know that e.g. $50\%$ of time a slave synchronizes
with a master, we still do not know much about the nature of these
events.  Thus, we shall next investigate in more detail for each of
the above four characteristic situations the representative time
series.

\begin{figure}[htb]
\centerline{\includegraphics{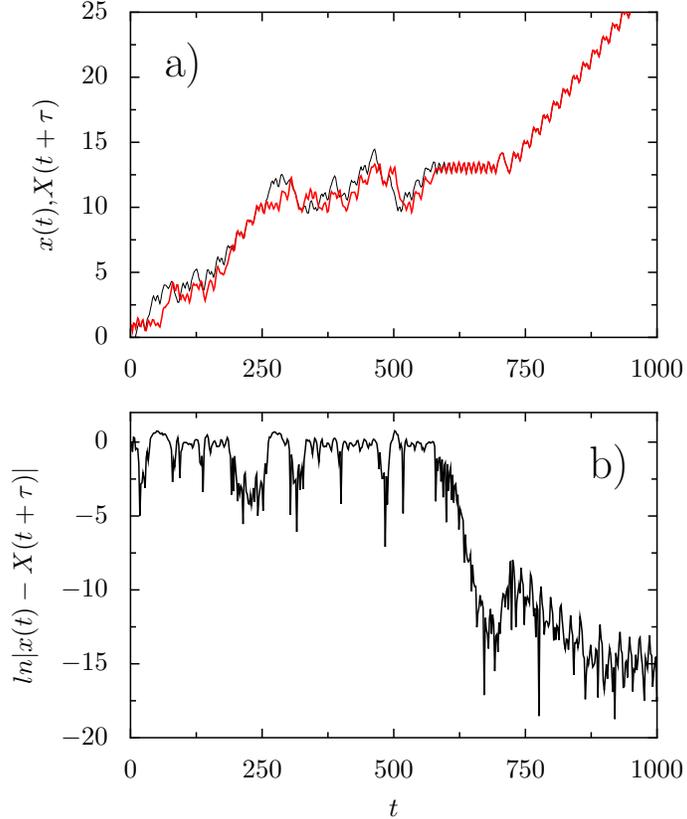}}
\caption{(Color online) Regular regime at the driving amplitude
  $a=0.074$.  (a): Trajectories for the master (black line) and slave
  (red line) ratchet dynamics.  (b): The logarithm of the absolute
  value of the difference between the future position of the master
  and the present one of the slave.  In the depicted scenario the
  slave already synchronizes to the transient of the master; later
  both systems together reach the growing, regular (period-2) orbit,
  see text.  The parameters are: $a = 0.074$, $b = 0.1$, $\omega=
  0.67$, $K = 0.2$, $\tau = 1.8$.}
\label{fig4}
\end{figure}

At the driving strength $a=0.074$, the master possesses a stable
regular trajectory, being characterized by a period-two orbit in the
corresponding Poincar\'e section, see e.g. Fig. 2 in \cite{ma1}).
Starting with independent random initial conditions within the same
period of the potential for the master and the slave, the master
reaches the stable orbit after a transient time. This is depicted in
Fig.  \ref{fig4}.  Already before the master has reached its stable
orbit the slave starts to synchronize at $t=600$. Only at $t=750$ both
the master and the slave reach the regular orbit and the
synchronization becomes even better, note the drop of the value
$\ln|x(t)-X(t+\tau)|$.  In this case $p=1$ , the slave will never
de-synchronize from the master.  Yet another scenario is possible:
depending on the chosen initial conditions, the slave and the master
may reach their regular orbits while keeping a spatial separation
equal to one period of the potential (which in our case is $1.0$).
Apparently, the attractor is strong enough to dominate the coupling
term $K\simeq 1.0$ and the system, although not synchronized, evolves
periodically (modulo the spatial period of the potential).  This
mechanism is responsible for the irregular shape of the dark area in
Fig.  \ref{fig3}.

\begin{figure}[htb]
\centerline{\includegraphics{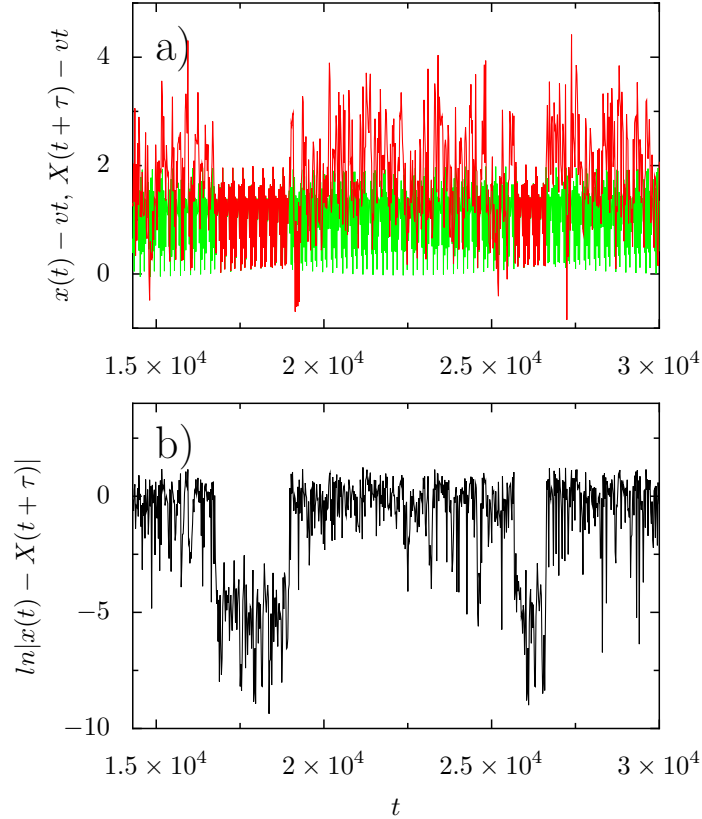}}
\caption{(Color online) Regular regime for a driving amplitude
  $a=0.081$.  (a): Trajectories for the master (black line) and slave
  (red line) ratchets (b): The logarithm of the absolute value of the
  difference between the respective positions.  In this case the
  master has already reached the regular orbit. The slave occasionally
  synchronizes and desynchronizes again.  The parameters are: $a =
  0.081$, $b = 0.1$, $\omega= 0.67$, $K = 0.0981$, $\tau = 3.4$.  The
  symbol $v$ stands for the average velocity of the ratchet dynamics.
  The trajectories are depicted in the frame moving with velocity $v
  \simeq -0.025$ in order to pronounce the oscillations. }
\label{fig5}
\end{figure}

At $a=0.081$, as in the previous case, the attractor of the master
exhibits a stable regular trajectory possessing a periodic orbit in
the Poincar\'e section (with period four \cite{ma1}).  There exists,
however, a dramatic difference in the synchronization behavior. In
Fig.  \ref{fig3} one can detect the appearance of an additional, well
synchronized region for small coupling strengths $K$, which spreads to
relatively large values of $\tau$. Also the temporal dynamics exhibits
another behavior. Let us inspect more closely the system dynamics for
the parameters $K=0.0981$ and $\tau=3.4$.  In Fig.  \ref{fig5} we
present a small portion of the trajectory at large times.  First, one
can observe that the master system has already reached the period-4
orbit. The slave, however, in contrast to the behavior in the previous
case, synchronizes and de-synchronizes in an intermittent manner.
Another observation is that the ``distance'' parameter
$\ln|X(t+\tau)-x(t)|$ is around $-5$ when synchronization takes place,
compared to a typical values $-15$ characterizing synchronization in
other parameter regimes. We also checked the behavior of the system at
$K=0.25$ and $\tau=1.6$, i.e.  where synchronization is observed for
all parameter values shown in Fig.  \ref{fig3}.  In that case the
scenario resembles the one at $a=0.074$, the slave and the master
reach the regular orbit and stay there forever.

\begin{figure}[htb]
\centerline{\includegraphics{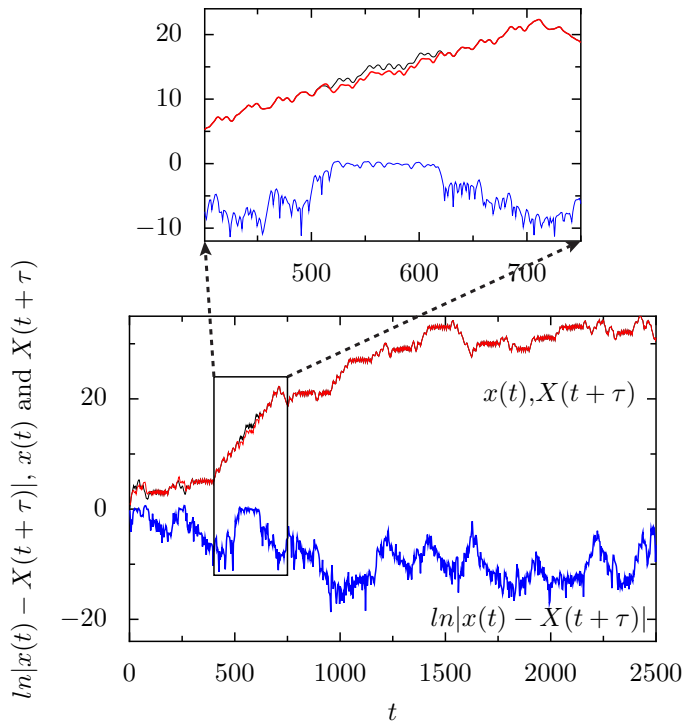}}
\caption{(Color online) Directed transport in the chaotic regime with
  a driving strength of $a=0.08$.  Depicted are the trajectories for
  the master and slave (upper curves, black and red lines,
  respectively) and the logarithm of the absolute value of the
  difference between the two positions (lower curves).  The slave
  occasionally synchronizes with, and subsequently de-synchronizes
  again from the master. The distance $\ln|X(t+\tau)-x(t)|$ exhibits a
  random walk like pattern.  The magnification in the upper panel
  depicts a short desynchronization event at $t=500$.  The parameters
  are: $a = 0.08$, $b = 0.1$, $\omega= 0.67$, $K = 0.305$, $\tau =   1.6$.}
\label{fig6}
\end{figure}

A typical chaotic transporting trajectory (possessing a small,
positive-valued transport-velocity) at the amplitude strength of
$a=0.08$ is depicted with Fig. \ref{fig6}. Similarly to the case
shown in Fig. \ref{fig5} we notice bursts of de-synchronization (or
synchronization, respectively, depending on the chosen parameters).
The ``distance'' parameter exhibits a random walk pattern, of the type
discussed in chapter 13 of Ref. \cite{piko}.

\begin{figure}[htb]
\centerline{\includegraphics{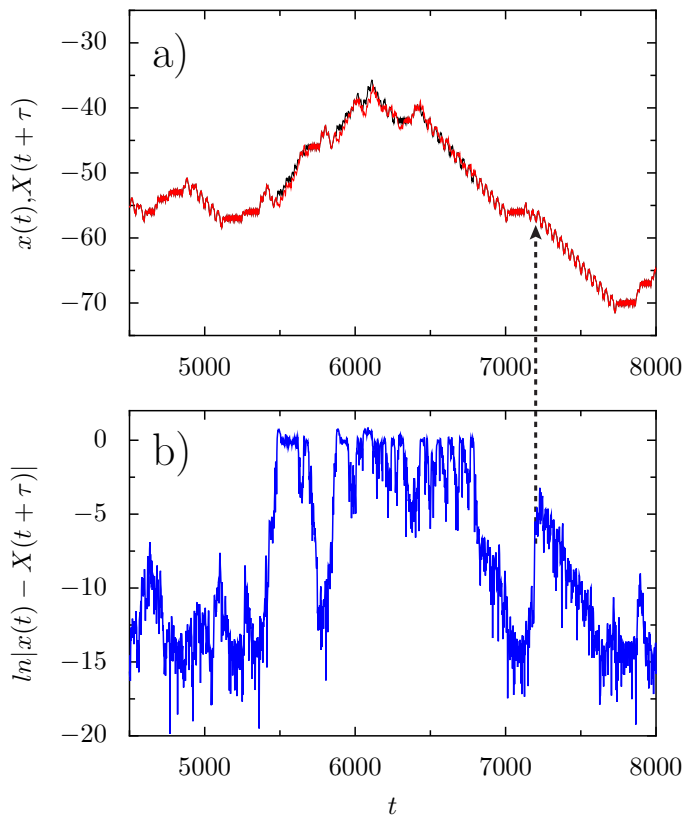}}
\caption{(Color online) Intermittent transport regime for the driving
  strength $a=0.0823$.  (a): Trajectories for the master (black line)
  and the slave (red line). (b): The logarithm of the absolute value
  of the difference between the two positions. The slave occasionally
  de-synchronizes from the master.  The parameters are: $a = 0.0823$,
  $b = 0.1$, $\omega= 0.67$, $K = 0.305$, $\tau = 1.6$.}
\label{fig7}
\end{figure}

The directed, transporting ratchet dynamics of the driven ratchet at
the driving strength $a=0.0823$ (see Fig. 2 in \cite{ma1}) is
intermittent. The region of synchronization (Fig.  \ref{fig3}) does
not significantly differ from the case at $a=0.08$. The trajectory
(see Fig.  \ref{fig7}), however, exhibits typical features of the
intermittency: the regular behavior is intermittently interrupted by
finite ``bursts'' in which the orbit behaves in a chaotic manner
\cite{ma1}.  Similarly to the chaotic case the slave occasionally
synchronizes with the master, and subsequently de-synchronizes again
from the master.  Those (de)synchronization events are not directly
connected with the ``bursts'' of the master dynamics!  However, we
have observed some regularities. In Fig. \ref{fig7} at $t=7200$ the
master changes its dynamics from a neighborhood of a period-two orbit
to the period-four orbit.  Simultaneously, the value of
$\ln|X(t+\tau)-x(t)|$ rises by ten orders of magnitude.  For this
particular event, however, the slave does not loose its
synchronization with the master.

\section{RESUME}

In summary, we numerically studied two deterministic ratchets coupled
unidirectionally via a time delay.  We established the conditions
under which one can obtain anticipated synchronization for the two
coupled transporting ratchet trajectories and postulated a necessary
stability criterion for the motion of the slave which is perfectly
confirmed by our numerical results. A further necessary condition for
the occurrence of synchronization is a strict relation between the
phases of the driving forces of the master and the slave.

Within the stable parameter region $K \tau < \pi/2$, we quantified the
degree of synchronization by means of its relative frequency $p$. For
$p$ values close to unity the slave ratchet anticipates the dynamics
of the master in an almost perfect way irrespectively of whether it
moves regularly or performs an intermittent or fully chaotic motion.
These results allow one to predict the directed transport features of
particles on a ratchet potential using a copy of the same system that
acts as a slave.

\acknowledgments

JLM gratefully acknowledges financial support from the Alexander von
Humboldt Foundation and UNAM through project DGAPA-IN-111000. PH 
acknowledges the  support by the Deutsche Forschungsgemeinschaft  via
project  HA1517/13-4.

\end{document}